\input psfig
\documentstyle [12pt]{article}

\begin{document}
\date{June 14, 1994}
\title{Nonperturbative Renormalon Structure of Infrared Unstable Theories}  
\author{A. Duncan and S. Pernice \\
Department of Physics and Astronomy \\
University of Pittsburgh, Pittsburgh, PA 15620\\}
\maketitle

e-mail: tony@dectony.phyast.pitt.edu, sergio@charm.pas.rochester.edu

\newpage

\begin{abstract}
The properties of a generalized version of the Borel Transform in infrared
unstable theories with dynamical mass generation are studied. The reconstruction of the nonperturbative structure is unambiguous in this
version. Various methods for extracting the singularity structure 
of the Borel Transform for lattice formulations of such theories are
explored, and illustrated explicitly with the O(N) sigma model. The
status of the first infrared renormalon in QCD is discussed. The
feasibility of a proposed technique for analytically continuing from
the left hand Borel plane (where nonperturbative information is available
via simulation of lattice field theory) to the positive real axis is
examined using the sigma model. 
\end{abstract}

\newpage
\section{INTRODUCTION}

  The problem of divergence of (renormalized) perturbation theory, which can be traced
back to the seminal work of Dyson in quantum electrodynamics \cite{Dyson}, is now 
recognized to lie at the core of any precise analytic understanding of the nonperturbative
structure of field theory. The need to face this problem squarely is particularly 
apparent in massless quantum chromodynamics (QCD), where the underlying dynamics is
specified by a single parameter $N$ (for gauge group $SU(N)$). Dimensionless ratios of
hadron masses in this theory depend only on $N$, which is of order unity ($N=$3) in the
physical world. Apart from the technically intractable 1/$N$ expansion, there  is no
natural {\em intrinsic} expansion parameter for the spectral and other low-energy
properties of the theory. If one insists on computing such quantities in a conventional
weak-coupling expansion, they either vanish formally to all orders, or yield asymptotic
expansions in terms of a running coupling which is naturally of order unity. Such expansions
are quantitatively useless in the absence of a reliable ``resummation" procedure.

  Certain superrenormalizable field-theories (e.g $\phi^{4}_{1,2,3}$ in the unbroken
phase) have been shown rigorously \cite{BenderWu},\cite{BorelField} to possess
Borel-summable perturbative expansions, so that the full content of the theory
(at least, information-theoretically) is exhausted by perturbation theory. The
spontaneously broken phases of these theories is typically not Borel-summable, but
even here an optimized reorganization of the perturbation theory can be shown
rigorously \cite{DuncanJones} to converge to the exact partition function, provide
the theory is formulated at finite volume. The failure of Borel-summability is far
more severe in QCD \cite{tHooft}, and no systematic analytic procedure is known,
even in principle, whereby the full nonperturbative structure of the theory could be
obtained on the basis of purely perturbative information.

  The divergence of the perturbation theory in QCD is far more than a merely technical
embarassment, as has recently been emphasized by Mueller \cite{Mueller}. It fundamentally
limits our ability to reliably compute important corrections (such as higher twist 
effects) to the vast phenomenology of high energy processes described by perturbative
QCD. Specifically, in those cases where one attempts to weld analytic perturbative
with numerical or phenomenological nonperturbative estimates for the same process, the
precision of the result is clouded by unavoidable resummation ambiguities on the
perturbative side. Consider, for example, a renormalization group controlled amplitude
$\Pi(Q^{2})$ in QCD, expressible as a formal (divergent asymptotic) series in the
running coupling $\alpha(Q^{2})$:
\begin{equation}\label{eq:piseries}
 \Pi(Q^{2}) \simeq \sum_{n} c_{n}\alpha^{n}(Q^{2})
\end{equation}
Quite generically, one may identify subsets of graphs in perturbation theory which 
contribute factorial growth to the dimensionless coefficients $ c_{n}$ at large $n$
\begin{equation}\label{eq:coeffn}
  c_{n} \simeq (\frac{b_{0}}{p})^{n} n^{\gamma} n!
\end{equation}
where $b_{0},b_{1}\equiv\frac{b_{0}^{2}\gamma}{2}$ are the first two coefficients of the
beta function, and $p$ is a positive integer. Using Stirling's approximation, the 
$n$-th term in the asymptotic series (\ref{eq:piseries}) may be written
\begin{equation}\label{eq:largenc}
 n!(\frac{b_{0}}{p})^{n}n^{\gamma}\alpha(Q^{2})^{n} \simeq e^{n\ln{(nb_{0}\alpha/p)}-n}
\end{equation}
which is minimal at $n_{\rm min}\simeq \frac{p}{b_{0}\alpha(Q^{2})}\simeq p\ln{\frac{Q^{2}}
{\Lambda^{2}}}$. The error in an asymptotic expansion is typically {\em at least as large}
(it may of course be much larger !) as the smallest term, here
\begin{equation}\label{eq:aerror}
 e^{-n_{\rm min}} \simeq (\frac{\Lambda^{2}}{Q^{2}})^{p}
\end{equation}
which is power suppressed at large $Q$ in just the way we expect for a higher twist
term. The moral is clear: higher twist effects are not precisely calculable in the
absence of a reliable resummation procedure. The same arguments in fact apply in
a variety of other important situations: for heavy quark expansions, where $1/M^{p}$
effects ($M$=heavy quark mass) may be confused with resummation ambiguities arising
in a static-quark quantity evaluated via the renormalization group in terms of a
perturbative expansion in $\alpha(M)$, or in estimating analytically power corrections
to lattice quantities in the lattice spacing $a$.

  The conventional Borel transform approach (for a thorough review, see \cite{BorelRef})
attempts to reconstruct the full nonperturbative structure of the theory from knowledge of the behavior of weak-coupling perturbation theory at large orders. There are two serious
problems with such an approach in infrared unstable theories like QCD. First, it is becoming
increasingly apparent that the large-order behavior is exceedingly complex \cite{Vainshtein}.
It is very difficult, even restricting oneself to a strictly limited subset of graphs,
to precisely identify the terms which give the truly dominant behavior at large order,
due to the subtle interplay of combinatoric and kinematic effects. Secondly, in a 
non-Borel-summable theory (and any infrared unstable theory with dynamical mass
generation falls into this category) there is in any case no reliable reconstruction
theorem based on the usual perturbative Borel transform, which develops singularities
(renormalons)
on the positive real axis of the Borel variable $s$, i.e on the Borel integral contour.
Fortunately, an alternative formulation of the Borel transform exists, first discussed
in QCD by t'Hooft \cite{tHooft}, and later by Crutchfield \cite{Crutchfield} and
David \cite{David}. This form of the transform is closely connected to the ``naive"
perturbative definition in Borel-summable theories, but can be used to give a precise
reconstruction theorem in a much wider class of theories. The singularity structure of
this transform is unambiguously related to the exact nonperturbative structure of the
full theory. But by the same token, this structure is not directly accessible in
conventional perturbation theory, but must be studied in an explicitly nonperturbative way.

  The primary objective of this paper is to suggest that useful information about the 
renormalon structure (i.e positive $s$ singularities)
 in infrared unstable theories can indeed be obtained using 
nonperturbative techniques (such as the large N expansion, or lattice theory).
 In Section 2, we review very quickly the conventional perturbative
Borel transform $\hat{B}(s)$, and explain its relation to the generalized transform
$B(s)$ and the density of states function $\tilde{B}(s) = {\rm disc} B(s)$. A
classification of renormalon singularities into those which are already present
at finite volume (``Type 1") and those which are only strictly speaking present
at infinite volume (``Type 2") is given. The
basic interconnections are further illustrated using spin models in 1 and 2
dimensions (the nonlinear sigma model) in Section 3. In Section 4 we explain 
the perturbative camouflage of renormalon singularity structure in disconnected
quantities such as the partition function (vacuum-vacuum amplitude). The status
of the first infrared renormalon singularity in QCD (recently raised by Brown,
Yaffe and Zhai \cite{BYZ}; see also \cite{BenekeZakh}) is discussed in
Section 4. It is shown that certain renormalon singularities of the full Borel
function $B(s)$ are immune to infrared cutoffs in ordinary perturbation theory
which remove the usual IR renormalons. Moreover, the presence of such a singularity
appears to be a generic feature once dynamical mass generation is assumed. In 
Section 5 we study the sigma model at large N on the lattice in order to 
establish the numerical feasibility of extracting renormalon singularities by
analytic continuation of nonperturbatively computed (say, by Monte Carlo simulation)
Euclidean amplitudes. Section 6 summarizes our conclusions and indicates directions
for further study.

\section{Borel Transform Technology}

\begin{figure}[htp]
\hbox to \hsize{\hss\psfig{figure=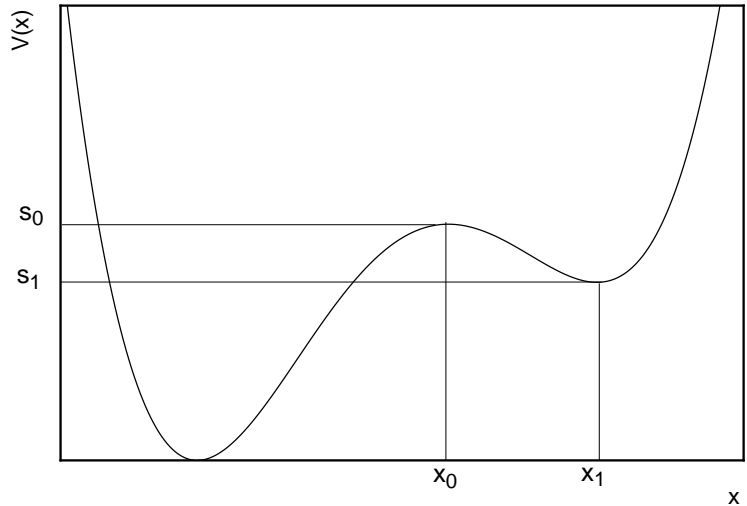,width=0.8\hsize}\hss}
\caption{NonGaussian Action leading to singular $\hat{B}(s)$}
\label{fig:V(x)}
\end{figure}

 The archetypal example of a Borel-summable perturbation theory is the nonGaussian 
integral
\begin{eqnarray}\label{eq:Ztoy}
  Z(f)&\equiv&\frac{1}{\sqrt(f)}\int_{0}^{+\infty}e^{-\frac{1}{f}V(x)}dx,\;\;\;V(x)=x^{2}+x^{4} \\
    &=&\int_{0}^{+\infty}e^{-x^{2}-fx^{4}}dx  \\
    &\simeq&\sum_{n=0}^{\infty} (-f)^{n}\frac{\Gamma(2n+\frac{1}{2})}{n!}\equiv\sum c_{n}f^{n+1}
\end{eqnarray}
Evidently the coefficients $c_n$ in the asymptotic expansion of $Z(f)$ for $f\rightarrow 0^{+}$
have the large order behavior $c_n \simeq (-1)^{n}4^{n}n!$ . Define the ``naive" Borel
transform
\begin{equation}\label{eq:naiveBor}
   \hat{B}(s) \equiv \sum_{n}\frac{c_n}{n!}s^{n}
\end{equation}
so that $Z_{0}(f)\equiv\int_{0}^{+\infty}e^{-s/f}\hat{B}(s)ds$ and $Z(f)$ have the same
asymptotic expansion for small $f$. In this case the analytic properties of $Z(f)$ are
such (see Ref \cite{BorelRef} for precise conditions) that $Z_{0}=Z$ and knowledge of
the ``naive" transform leads to a precise reconstruction of Z for all $f$. We shall 
actually be using a more powerful version of the Borel method in most of this paper,
which will circumvent the need for theorems of this type whcih certify the preceding
procedure. The shortcomings of $\hat{B}(s)$ for non-Borel-summable theories, where 
the sign oscillation in $c_n$ is absent in some uncancelled set of contributions at
large order ({\em not necessarily the dominant ones}!) are immediately apparent: 
$\hat{B}(s)$ develops singularities on the positive real $s$ axis, rendering the
reconstruction integral ambiguous (if the singularities are non-integrable). The most
straightforward way in which such singularities can arise is illustrated in Fig. \ref{fig:V(x)},
where the ``action" $V(x)$ possesses a secondary nonperturbative extremum.

To expose the singularity in $\hat{B}(s)$ write \cite{ZinnJustin}
\begin{eqnarray}\label{eq:ZBtilde}
  Z(f)&=&\int e^{-V(x)/f}dx \\
      &=&\int_{0}^{+\infty} ds e^{-s/f} \int dx \delta(s-V(x)) \\
      &=&\int_{0}^{+\infty} ds e^{-s/f} \tilde{B}(s)
\end{eqnarray}
where 
\begin{equation}\label{eq:Btildedef}
  \tilde{B}(s)\equiv\int dx\delta(s-V(x)) =\sum_{V(x_i)=s}\frac{1}{|V^{\prime}(x_{i})|}
\end{equation}
If $V$ has a nontrivial extremum where $V^{\prime}(x_i)=0,s_i =V(x_i)>0$, we will have
necessarily $B(s_i)=\infty$, a singularity (in this case, an integrable square-root
singularity, as $\tilde{B}\simeq 1/\sqrt{s-s_i}$) for positive $s$. To the extent
that the interchange of integrations carried out in Eq(\ref{eq:ZBtilde}) is legal, 
the reconstruction of $Z$ from $\tilde{B}$ must be precise. In fact $\tilde{B}$
is intimately related to the more general Borel function to be introduced below.
In general, however, $\hat{B} \neq \tilde{B}$! Of course, the asymptotic expansions
for small $s$ of $\hat{B},\tilde{B}$, which determine the asymptotic expansion of $Z(f)$
for small $f$, must agree. As pointed out originally by t'Hooft \cite{tHooft}, the existence of
Euclidean extrema of the action (instantons) in QCD implies in a similar way the 
presence of singularities in $\tilde{B}(s)$ at $s=S_{\rm inst}$. Moreover, the location
of the singularities (though not their strength) is universal (see (\cite{tHooft}):
once a singularity appears
in any amplitude of the theory, it is expected to propagate to all others. 

 For a class of non-Borel-summable theories (the archetype is the double well anharmonic
oscillator) an optimized reorganization of perturbation theory can be shown to converge with
exponential rapidity \cite{DuncanJones} to the exact partition function. These convergence
proofs apply only when the theory is formulated at finite volume, however, limiting the
practical usefulness of such methods in higher dimensional field theories. The approach 
taken here will be to employ a generalized Borel Transform which allows a ``bulletproof"
reconstruction of the full theory, avoids resummation ambiguities, and is more closely
tied to the path integral formulation of the theory \cite{Crutchfield}. Our ultimate
objective is to explore the potential of the presently available nonperturbative lattice
techniques for computing the singularity structure of this transform.

  Consider a typical infrared-unstable (asymptotically free) theory, such as the O(N) 
sigma model in 2 dimensions, or QCD in 4 dimensions, defined by a functional integral
\begin{equation}\label{eq:Zphi}
  Z(f)=\int {\cal D}\phi e^{-\frac{1}{f}S[ \phi ]}
\end{equation}
where $\phi$ is a generic set of fields, $S[ \phi])$ the Euclidean action, and $f$ a bare
coupling constant. A compact lattice formulation  ($\int{\cal D}\phi =1$)
of the theory is supposed, so that a
UV cutoff is always present, and we have logarithmic asymptotic freedom
\begin{equation}\label{eq:asymfree}
   f \sim \frac{1}{b_0 \ln{(\Lambda/m)}},\;\;\;\Lambda >> m
\end{equation}
where $m \sim$ (physical correlation length)$^{-2}$, $\Lambda \sim$ (short distance cutoff)$^{-2}$.
For both sigma models and QCD on a {\em finite} lattice, $0\leq S[ \phi]\leq S_{\rm max}$ is a
bounded continuous function of the fields, and the integral $\int {\cal D}\phi$ is compact.
Define the generalized Borel Transform $B(s)$ as the Laplace transform with respect to
inverse coupling, as follows
\begin{equation}\label{eq:GBT}
  B(s) \equiv \int_{0}^{+\infty} e^{s/f}Z(f)d(\frac{1}{f}),\;\;\;{\rm Re(}s)<0
\end{equation}
The integral is convergent at large $f$ for all $s$ (as $e^{s/f}Z(f)\rightarrow 1$) and for
Re($s$)$<0$ as $f\rightarrow 0^{+}$. Thus $B(s)$ is left-half-plane analytic, with a 
cut on the positive real axis for $0<s<S_{\rm max}$. Indeed, inserting the functional
integral expression (\ref{eq:Zphi}) in (\ref{eq:GBT}), and defining $x\equiv\frac{1}{f}$,
\begin{eqnarray}\label{eq:GBint}
  B(s) &=& \int{\cal D}\phi \int_{0}^{+\infty} e^{x(s-S[ \phi])}dx  \\
       &=& \int{\cal D}\phi \frac{1}{S[ \phi]-s}
\end{eqnarray}
which is manifestly analytic in the $s$-plane cut along the positive real axis between
$s=0$ and $s=S_{\rm max}$. The discontinuity of $B(s)$
across this cut is
\begin{equation}\label{eq:disc}
  {\rm Im} B(s=s_{R}+i\epsilon)=\pi\int{\cal D}\phi\delta(s_{R}-S[ \phi]) \equiv \pi\tilde{B}
(s_{R})
\end{equation}
The physical interpretation of this discontinuity is clear: $\tilde{B}(s)$ is the
density of configurations in the functional integral with action equal to $s$
(cf. Eq(\ref{eq:Btildedef})). Singularities of $\tilde{B}$ are expected
wherever there is a sharp local enhancement in the number of configurations, for
example wherever the action has a local extremum. 
 
 The rigorous reconstruction of $Z(f)$ from $\tilde{B}$ is trivial in this framework.
The vertical contour of the inverse Laplace transform may be wrapped around the cut
on the positive $s$ axis to give an integral over the discontinuity of $B(s)$, i.e.
over $\tilde{B}$. More directly, we may simply observe that it follows from
Eq(\ref{eq:Zphi},\ref{eq:disc}) that
\begin{equation}\label{eq:fullreconstr}
  Z(f)=\int_{0}^{+\infty}\tilde{B}(s)e^{-s/f} ds
\end{equation}
This reconstruction is exact: there are no routing ambiguities in the $s$-integral.
Of course, $\tilde{B}$ is in general a distribution (containing potentially $\delta$
singularities), not obtainable purely from knowledge of weak-coupling perturbation
theory. Instead, it must be computed nonperturbatively by analytic continuation of
$B(s)$, computed nonperturbatively in the region Re($s)<0$, where the defining integral
(\ref{eq:GBT}) exists. It should be noted at this point that $\tilde{B}(s)$, in
contrast to $B(s)$ {\em may have singularities} in the left-half-plane (corresponding
to non-principal-sheet singularities of $B(s)$): the {\em ultraviolet} renormalons of
the renormalized perturbation theory in QCD are of this type.

 It is convenient to distinguish between two types of singularities appearing in 
$\tilde{B}(s)$. Exact local extrema of the Euclidean action may exist even for
systems with a finite number of degrees of freedom (e.g. field theories 
formulated on a finite lattice). Such extrema give rise to actual singularities
(henceforth called ``Type 1") of $\tilde{B}$ even at finite volume, when the
theory is cutoff in the infrared. For example, one can find lattice analogs of the
instanton solutions of the classical 2D O(3) sigma model using a simple numerical
annealing procedure to solve the lattice Euclidean field equations. Other 
singularities of $\tilde{B}$ only appear at infinite volume, and will be called 
Type 2 singularities. The famous infrared renormalons of large order perturbation
theory are connected to singularities of this type. As numerical simulations of
nonperturbative behavior are necessarily restricted to finite volume lattice
theories, the question arises whether singularities of this type are really
accessible using simulation techniques. We shall see below that the question is
complicated by the nonuniformity of the analytic continuation to positive $s$
in the large volume limit, which will require a modification of the definition 
(\ref{eq:GBT}) in the finite volume situation. For the time being, let us 
illustrate the absence of a true singularity in (\ref{eq:GBT}) in a discrete
system with a simple toy model. Consider a mass gap relation connecting a bare
coupling $f$ and a dynamically generated mass $m$ of the form
\begin{equation}\label{eq:gapL}
  \frac{1}{f}=\frac{1}{L}\sum_{n=0}^{L}\frac{1}{m+n/L} \rightarrow \ln{(\frac{1+m}{m})},
L\rightarrow\infty
\end{equation}
The ``infinite volume limit" thus gives $m=\frac{e^{-1/f}}{1-e^{-1/f}}$, and Borel 
transforms $B(s)$ of any analytic function $G(m)$ expandable at $m=0$ will display
simple poles at $s=0,1,2,3...$ For example, if $G(m)=\frac{1}{1+m}$, we have
(at $L=\infty$) $B(s)=-1/s-1/(1-s)$ (for other choices of the ``momentum" in the
denominator of $G(m)$, there will typically be poles at all integer $s$).
 On the other hand, for finite $L$, we have
\begin{eqnarray}\label{eq:BLS}
  B_{L}(s)&=&\int_{0}^{\infty}e^{s/f}\frac{1}{1+m}d(\frac{1}{f})  \nonumber \\
          &=&\int_{0}^{\infty}\frac{1}{L}\sum\frac{1}{(m+n/L)^{2}}\frac{1}{1+m}
e^{s\cdot\frac{1}{L}\sum\frac{1}{m+n/L}}dm
\end{eqnarray}
where the $m$-integral in (\ref{eq:BLS}) is convergent for Re($s)<0$. We may analytically
continue to $s=1$ by deforming the $m$ contour to avoid a singularity from the $n=0$
term in the exponent,i.e. by keeping arg($\frac{s}{m}$) fixed near $m=0$. This can
be done without encountering any pinches of the integration contour. Consequently,
$B_{L}(s=1)$ is perfectly finite for finite $L$. In Section 6 we shall see that $B_{L}(s)$
needs to be modified so that a smooth approach to the infinite volume behavior for
positive $s$ is obtained. 

\section{Non-Borel-summable spin models}
\subsection{Separable models (D=1)}

\begin{figure}[htp]
\hbox to \hsize{\hss\psfig{figure=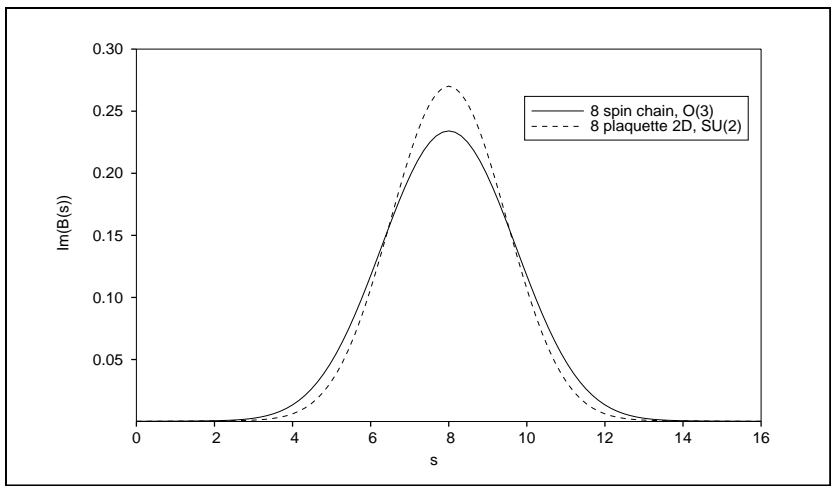,width=0.8\hsize}\hss}
\caption{Borel Transform $\tilde{B}$ for $L=8$ spin chain}   
\label{fig:L8}
\end{figure}

 A simple model which already serves to illustrate several features of the
renormalon singularity structure in more general non-Borel theories is the
1D spin chain defined by the action
\begin{equation}\label{eq:spinchain}
   S=\sum_{i=1}^{L}(1-\hat{\phi}_{i}\cdot\hat{\phi}_{i+1})
\end{equation}
Here the $\hat{\phi}_{i}$ are unit 3-vectors and free boundary conditions
are chosen so that the model factorizes in a trivial way. Indeed, the
partition function is
\begin{eqnarray}\label{eq:spinpart}
   Z_{L}(f)&=&\int\Pi_{i}d\hat{\phi}_i e^{-S[\hat{\phi}]/f} \nonumber \\
           &=& (\frac{f}{2})^{L}(1-e^{-2/f})^{L}
\end{eqnarray}
showing a trivial separation of the ``perturbative" ($\propto f^{L}$) and
``nonperturbative" ($\propto f^{L}e^{-2n/f}$) terms. For $L=1$, the
Borel Transform of $Z$ is 
\begin{equation}\label{eq:B1s}
   B_{1}(s)=\frac{1}{2}\ln{(\frac{2-s}{-s})}
\end{equation}
with discontinuity
\begin{equation}\label{eq:B1tilde}
  \tilde{B}_{1}(s)\equiv \frac{1}{\pi}{\rm Im}\;B(s+i\epsilon)=\frac{1}{2}\theta(s)
\theta(2-s)
\end{equation}
Note that  poles only appear in  $\tilde{B}_{1}(s)$ after a derivative with respect
to $s$. This effectively (cf. Eq(\ref{eq:GBT})) removes the overall perturbative
factor of $f$ (associated with the measure in the path integral), exposing the
nonperturbative behavior. For $L$ large, $\tilde{B}_{L}(s)$ is even smoother (each
prefactor of $f$ effectively integrates with respect to $s$). The singularity
structure is numerically invisible in this case. In Fig. \ref{fig:L8} we show
$\tilde{B}_{8}(s)$. It is clearly difficult to tell that there are actually
singularities at $s=2,4,6,8,...$! In Section 4 we shall see that this perturbative 
camouflaging of the singularity structure is typical of disconnected quantities
like $Z$, but is not a problem when connected quantities (with a well-defined 
infinite volume limit) are considered.

 In agreement with the usual analogy between spin models in D and gauge models
in 2D dimensions, the separable model of 2D pure lattice gauge theory is
found to have an essentially identical Borel structure. If we take the gauge
group to be SU(2), the partition function for a system of $L$ plaquettes (with
free boundary conditions) is found to be $Z_{L}(f)=[2fe^{-1/f}I_{1}(\frac{1}{f})]^{L}$.
For $L=1$, the Borel Transform is $\tilde{B}_{1}(s)=\sqrt{s(2-s)}$. The Borel
discontinuity (=density of states in functional integral) for a system of
8 plaquettes is also plotted in Fig(\ref{fig:L8}) for comparison with the
spin chain case. Once again, the singularities at $s=2,4,6,...$ are numerically
invisible.

\subsection{Borel Structure of 2D Sigma model}

  The partition function of the continuum 2D Sigma model is given by
the Euclidean functional integral
\begin{equation}\label{eq:2DZ}
 Z(f)=\int {\cal D}\hat{\phi}(x) e^{-\frac{1}{2f}\int d^{2}x|\partial_{\mu}
\hat{\phi}|^{2}}
\end{equation}
 Here $\hat{\phi}$ is a unit N-vector field, $f$ is a bare coupling, and a
UV cutoff is presumed. On a finite 2 dimensional $L$x$L$ lattice, the
corresponding quantity is
\begin{eqnarray}\label{eq:2DLatt}
  Z_{L}(f)&=&\int\Pi_{n}d\vec{\phi}_{n}d\rho_{n}e^{i\sum_{n}\rho_{n}
(\vec{\phi}_{n}^{2}-1)-\frac{N}{2f}\sum_{n,\mu}|\Delta_{\mu}
\vec{\phi}_{n}|^{2}}  \nonumber \\
      &=&(\frac{f}{N})^{NL^{2}}\int\Pi_{n}d\vec{\phi}_{n}d\rho_{n}e^{i\sum_{n}\rho_{n}
(\vec{\phi}_{n}^{2}-N/f)-\frac{1}{2}\sum_{n,\mu}|\Delta_{\mu}
\vec{\phi}_{n}|^{2}}              
\end{eqnarray}
Here $n$ labels lattice sites, and an auxiliary condensate field
$\rho_{n}$ has been introduced to implement the nonlinear constraint.
Integrating out the $\vec{\phi}$ fields
\begin{equation}\label{eq:Zrho}
  Z_{L}(f)=(\frac{f}{N})^{NL^{2}}\int\Pi_{n}d\rho_{n} e^{-\frac{N}{2}
S_{\rm eff}[\rho_n]}
\end{equation}
where
\begin{equation}\label{eq:Seff}
  S_{\rm eff}= {\rm TrLn}(-\Delta^{2}-2i\rho)+\frac{2i}{f}\sum_{n}\rho_n
\end{equation}
For large $N$, the functional integral in Eq(\ref{eq:Zrho}) is dominated
by a translationally invariant complex saddle point at $\rho_{n}=\rho_{\rm sad}$
where
\begin{equation}\label{eq:gapeq}
  \frac{1}{f}=\frac{1}{L^{2}}\sum_{\vec{k}}\frac{1}{d(\vec{k})+m},\;\;m\equiv
-2i\rho_{\rm sad}
\end{equation}
Here the kinematic momentum factor appearing in the propagator denominator
is $d(\vec{k})\equiv 4(\sin{(\pi k_x/L)}^{2}+\sin{(\pi k_y/L)}^{2})$. Thus the
$1/N$ expansion \cite{1overN} generates immediately a dynamical
(squared) mass $m(f)$. Note that for any finite $L$, the dynamical mass
$m(f)$ defined implicitly in Eq(\ref{eq:gapeq}) is actually {\em analytic}
at $f=0$. Any analytic function of $m$ (e.g. the large $N$ limit of the
$\phi$ propagator at momentum $\vec{k}$, namely $\frac{1}{d(\vec{k})+m}$)
will therefore have a conventional Borel transform $\hat{B}(s)$ which is 
entire in $s$. The full Borel transform $B(s)$ of the large $N$ propagator
has a cut on the positive axis, but no true singularities of $\tilde{B}$
for finite $L$. The problem of recovering the singularities of the large
volume limit from lattice data will be discussed in Section 6. Type 1
(instanton) singularities which would surface already at finite $L$ are
of course absent in the large $N$ limit.

  The large $N$ result for the partition function is
\begin{equation}\label{eq:ZLN}
 Z_{L} \simeq (f/N)^{NL^{2}}e^{-\frac{NL^{2}}{2}(\frac{1}{L^{2}}\sum_{\vec{k}}
\ln{(d(\vec{k})+m)}-m/f)}
\end{equation}
Again, as for the spin chain case, the perturbative prefactor will 
numerically camouflage the nonperturbative structure. 

  For future reference, we list here several important features of the large $N$ 
limit in the {\em continuum} theory. Initially, assume both infrared and 
ultraviolet cutoffs to be present for the modes of the $\vec{\phi}$ field:
$\lambda < k^{2} < \Lambda$. The mass gap equation then becomes
\begin{eqnarray}\label{eq:gapcont}
   \frac{1}{f}&=&\int\frac{d^{2}k}{(2\pi)^{2}}\frac{1}{k^{2}+m} \nonumber \\
              &=&\frac{1}{4\pi}\ln{(\frac{\Lambda +m}{\lambda +m})}
\end{eqnarray}
giving a dynamical squared mass
\begin{equation}\label{eq:masscut}
  m=\frac{\Lambda e^{-4\pi/f}-\lambda}{1-e^{-4\pi/f}}
\end{equation}
displaying the characteristic essential singularity at $f=0$. The singularity is
present even though $\lambda \neq 0$ (and conventional perturbative IR
renormalons are absent:cf. Section 5)! The large $N$ limit of the $\vec{\phi}$
propagator in momentum space 
\begin{equation}\label{eq:phiprop}
\frac{1}{q^{2}+m}=\frac{1}{q^{2}}-\frac{m}{q^{4}}+\frac{m^{2}}{q^{6}}-..
\end{equation}
possesses a meromorphic Borel transform $B(s)$. Taking $\lambda=0$ for simplicity
and changing variables to $x\equiv 1/f$
\begin{equation}\label{eq:Bsprop}
\int_{0}^{+\infty}dx e^{sx}(\frac{\Lambda e^{-4\pi x}}{1-e^{-4\pi x}})^{n}
=\Lambda^{n}(\frac{1}{4\pi n-s}+\frac{n}{4\pi(n+1)-s}+..)
\end{equation}
so there are simple poles at $s=4\pi,8\pi,12\pi,...$. Switching on the infrared
cutoff $\lambda$ modifies the residue, but not the location of these poles.
This apparently paradoxical situation is elucidated further in Section 5.

 The meromorphic character of the Borel transform of the propagator extends to other
Green's functions in the large $N$ limit. The two-current correlator, for example,
is defined by
\begin{eqnarray}\label{eq:2curr}
\Pi^{\mu\nu}(q)&=&\frac{1}{N^{2}}\int d^{2}x<0|J^{\mu}_{ab}(x/2)J^{\nu}_{ab}(-x/2)|0>e^{-iq
\cdot x}  \nonumber \\
 J^{\mu}_{ab} &\equiv& \frac{1}{2}\phi_{a}\partial^{\mu}\phi_{b}
\end{eqnarray}
Neglecting terms of $O(q^{2}/\Lambda)$ and subtracting at $q=0$, one has (in the large N
limit)
\begin{equation}\label{eq:pimunu}
 \Pi^{\mu\nu}=-\frac{1}{8\pi}(g^{\mu\nu}-\frac{q^{\mu}q^{\nu}}{q^{2}})
\int_{0}^{1}dx\ln{(1+x(1-x)\frac{q^{2}}{m})}
\end{equation}
The Borel transform can be computed by leaving the Feynman parameter integration
to the end. Again neglecting terms of order $q^{2}/\Lambda$ (but {\em not} $m/q^{2}$!) 
the relation between $m$ and $f$ is effectively $m=\Lambda e^{-4\pi/f}$ so 
\begin{equation}
 B^{\mu\nu}(s)=-\frac{1}{32\pi^{2}}\Lambda^{s/4\pi}(g^{\mu\nu}-\frac{q^{\mu}q^{\nu}}{q^{2}})
\int_{0}^{1}dx\int_{0}^{\Lambda}dm\; m^{-1-s/4\pi}\ln{(1+\frac{x(1-x)q^{2}}{m})}
\end{equation}
We are interested in the singularity structure of $B^{\mu\nu}(s)$ for Re($s)>0$ (the generalized
Borel Transform is left-half-plane analytic). Since
\begin{equation}
 \int_{0}^{1}dx\int_{\Lambda}^{+\infty}dm\; m^{-1-s/4\pi}\ln{(1+\frac{x(1-x)q^{2}}{m})}
\end{equation}
is analytic for Re$(s)>0$, we may complete the $m$ integral to $\int_{0}^{+\infty}$ 
without altering the singularity structure in the right-half-plane. One then finds
\begin{equation}\label{eq:Bmunuans}
B^{\mu\nu}(s;q)\simeq\frac{1}{32\pi^{2}}(g^{\mu\nu}-\frac{q^{\mu}q^{\nu}}{q^{2}})
(\frac{\Lambda}{q^{2}})^{s/4\pi}\frac{\Gamma(s/4\pi)\Gamma(-s/4\pi)\Gamma(1-s/4\pi)^{2}}
{\Gamma(2-s/2\pi)}
\end{equation}
which has double poles at $s=4\pi,8\pi,...$ These poles would be visible as sharp peaks
if $B^{\mu\nu}(s)$ could be computed at Re$(s)<0$ and then analytically continued to
$s=s_{R}+i\gamma,\gamma <<s_{R}, s_{R}>0$. Note that $B^{\mu\nu}(s)$ satisfies the
Brown-Yaffe-Zhai relation \cite{BYZ}
\begin{equation}
{\rm disc} B^{\mu\nu}(s;q^{2}=-Q^{2})=\sin{(\pi b_{0}s)}B^{\mu\nu}(s;Q^{2}),\;\;b_0=\frac{1}{4\pi}
\end{equation}

  From the foregoing, it is apparent that the nonlinear sigma-model in the large $N$
limit exhibits an interesting and elaborate renormalon pole structure \cite{David} in the
right-half Borel plane. This is so even though a conventional perturbation expansion
applied to the large $N$ result leads to trivial results, insofar as the nonperturbative
structure enters in this limit entirely through the quantity $m\sim e^{-4\pi/f}$ which
vanishes to all orders in a formal expansion around $f=0$.

\section{Perturbative Camouflage of Renormalon Singularities}

\begin{figure}[htp]
\hbox to \hsize{\hss\psfig{figure=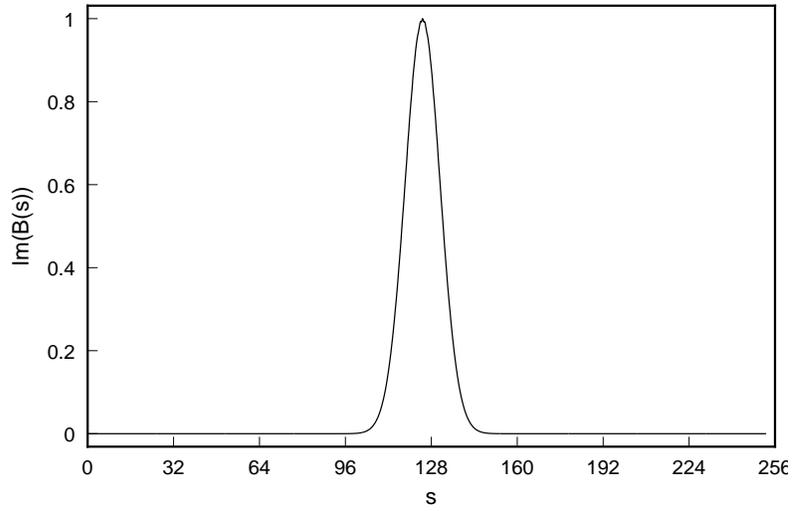,width=0.8\hsize}\hss}
\caption{Density of configurations for 2D O(3) Sigma model}   
\label{fig:2DBs}
\end{figure}

 Formally the most convenient field-theoretic quantity from the point of view of the
generalized Borel function $B(s)$ is the basic partition function $Z(f)$ of Eq(\ref{eq:Zphi}),
as the discontinuity of $B(s)$ in this case is nothing but the density of configurations
in a given action shell in the functional integral. Unfortunately, $Z$ does not possess
an infinite volume limit, and as we saw in Section 3, the Type 1 singularities of $B(s)$
for a system of large but finite volume $V$ are in any event camouflaged by 
perturbative prefactors $\sim f^{V}$ which smooth the $s$-dependence (as multiplication by
$f$ is equivalent to integration by $s$). The Borel transform of $Z(f)$ is
very accurately computable by Monte Carlo techniques: from the density of states
interpretation described in Section 2 one has only to histogram the frequency
of configurations
generated in a Monte Carlo simulation (relative to a conveniently chosen reference
action) versus action. The result of a simulation of
$\tilde{B}(s)$ for a 8x8 2D sigma model is shown in Fig(\ref{fig:2DBs}), to be compared
with Fig(\ref{fig:L8}) for the spin chain or 2D QCD. Note that  the singularities at positive $s$
are numerically invisible.

  The connected vacuum amplitude $W(f)\equiv\ln{Z}/V$ is clearly 
preferable to $Z$ in this regard. Apart from having a well-defined 
infinite volume limit, perturbative prefactors in $Z$ appear as
additive contributions in $W$, effectively uncovering the renormalon
structure of the theory. We may expect the same to be true of other
connected Green's functions (e.g. two-point functions) of the theory,
which are expressible as source-derivatives of $W$. For example,
consider the trivial spin-chain model of Section (3.1):
\begin{equation}\label{eq:WLspin}
  W_L =\frac{1}{L}\ln{Z_L}=\ln{(f/2)}+\ln{(1-e^{-2/f})}
\end{equation}
the Borel transform $B_W$ of which exhibits simple poles at
$s=2n,n=1,2,..$ with residue $1/n$. The continuation to positive $s$
thus exhibits Lorentzian peaks:
\begin{eqnarray}\label{eq:bwpoles}
  B_W(s)&\sim&\sum_n\frac{1}{n}\frac{1}{s-n}  \nonumber \\
  {\rm Im} B_W(s_R+i\gamma)&\sim& \gamma\sum\frac{1}{n}\frac{1}{(s_R-n)^{2}+
\gamma^{2}}
\end{eqnarray}

Assuming the Borel discontinuity $\tilde{B}(s)$ of $Z$ has been computed
(say, by Monte Carlo) sufficiently accurately, the analytic continuation
of $Z(f)$ to $x\equiv\frac{1}{f}=iy$ ($y$ positive real) is available as
the Fourier transform of $\tilde{B}$. Recall
\begin{equation}
  Z(x=1/f)=\int_{0}^{+\infty}ds \tilde{B}(s)e^{sx}
\end{equation}
while
\begin{equation}\label{eq:bwsdef}
B_W(s)\equiv\int_{0}^{+\infty}dx e^{sx}\ln{Z(x)},\;\;{\rm Re}(s)<0
\end{equation}
Analytically continue $B_W$ from real negative $s$ to $s=s_R+i\gamma$
($s_R,\gamma>0$) by simultaneously deforming the x-contour
in Eq(\ref{eq:bwsdef}) to $x=iy$. Thus
\begin{equation}
 {\rm Im}B_W(s_R+i\gamma)={\rm Re}\int_{0}^{\infty}dy e^{is_R y-\gamma y}
\ln{\int_{0}^{S_{\rm max}}ds \tilde{B}(s)e^{isy}}
\end{equation}

\begin{figure}[htp]
\hbox to \hsize{\hss\psfig{figure=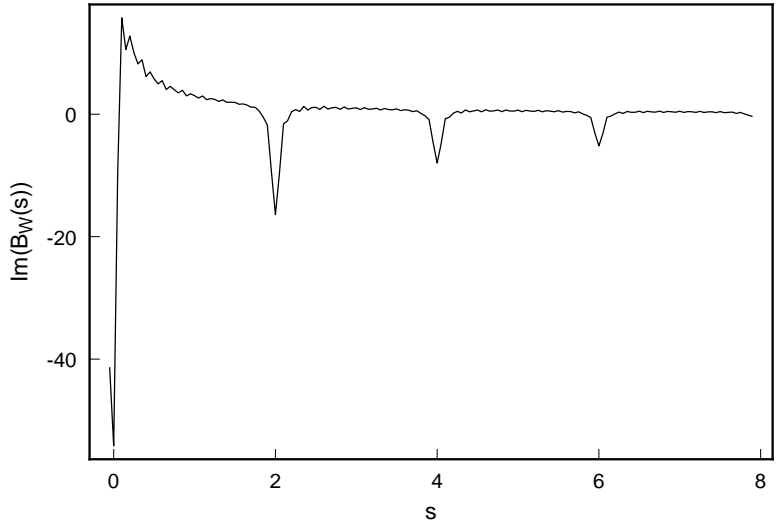,width=0.8\hsize}\hss}
\caption{Borel singularities of $W$ in 1D O(3) spin chain}    
\label{fig:2DBW}
\end{figure}

The result of this computation for the spin chain model is shown in 
Fig(\ref{fig:2DBW}) for $\gamma=0.05$. The singularities at $s=2,4,6,..$
are clearly visible. (The oscillatory structure visible in
Fig(\ref{fig:2DBW}) is due to the fact that the integral over
$y$, performed numerically, was cut off at the upper end).
It would be clearly of great interest to 
locate the Type 1 singularities in more interesting lattice field
theories along these lines. At present most of the discussion of
renormalon structure centers on correlation functions (e.g. the
Euclidean current-current correlator) and we shall focus instead on
an alternative approach which allows the extraction of renormalon
singularities (of both types) of connected Green's functions.

\section{Status of the First Infrared Renormalon in QCD}

\begin{figure}[htp]
\hbox to \hsize{\hss\psfig{figure=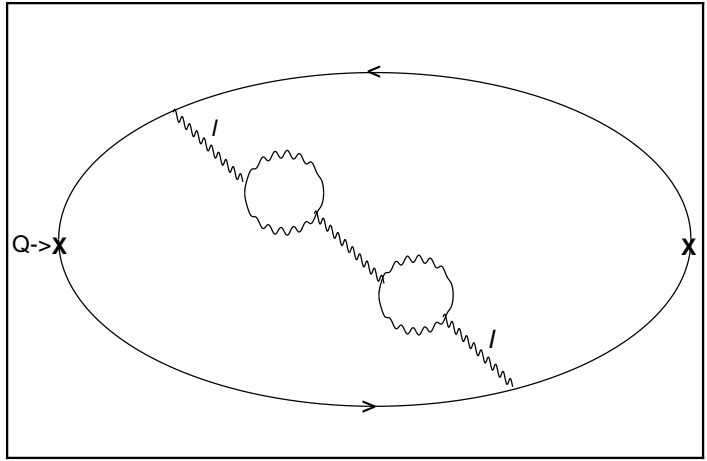,width=0.8\hsize}\hss}
\caption{One of a set of bubble graphs contributing to the leading
perturbative renormalon in QCD}
\label{fig:IRgraph}
\end{figure}

  It has been known for some time that evidence for a nontrivial
singularity structure in the right-half-plane of the Borel variable $s$
can be detected within the framework of conventional (weak-coupling)
perturbation theory. Consider an asymptotically free field theory
(such as massless QCD) with no intrinsic mass scale at the classical
level. Any scalar Euclidean momentum space correlation function
$\Pi(Q^{2})$ corresponding to a renormalization group invariant
quantity will have a formal perturbative expansion in powers of
the running coupling $\alpha(Q^{2})$:
\begin{equation}\label{eq:PiQ}
   \Pi(Q^{2})\sim \sum_{n}c_n \alpha^{n}(Q^{2})
\end{equation}
An example of great phenomenological interest is the hadronic
electromagnetic current correlator, whence $\Pi(Q^{2})=
\frac{1}{3Q^{2}}\int d^{4}x e^{iq\cdot x}<0|T[j^{\mu}(x)j_{\mu}(0)]|0>$.
Perturbative infrared renormalons are associated with subsets of 
graphs which result in a contribution to the coefficients $c_n$
at large $n$ of the form
\begin{equation}\label{eq:cnir}
  c_n \sim n!(\frac{b_0}{p})^{n}
\end{equation}
where $b_0=\frac{11-2N_f/3}{4\pi}$ is the first nontrivial 
coefficient in the beta-function and $p$ is an integer.
The behavior (\ref{eq:cnir}) implies a singularity in the
naive Borel Transform $\hat{B}(s)$ at $s=p/b_0$. Such terms
can be seen to arise from sets of graphs generating the leading
infrared logarithms in the current-current correlator
(see Fig(\ref{fig:IRgraph}). Introducing an infrared cutoff
$\lambda$, the contribution of the subset of graphs shown in
Fig(\ref{fig:IRgraph}) is found to be \cite{Mueller}
\begin{equation}\label{eq:IRint}
  \int_{\lambda^{2}} l^{2}dl^{2}\frac{\alpha(Q^{2})}{1
+b_0\alpha(Q^{2})\ln{(\frac{l^{2}}{Q^{2}})}}\sim
\sum\alpha^{n+1}(Q^{2})(b_0)^{n}\int_{c}^{\ln{(Q^{2}/\lambda^{2})}}
x^n e^{-2x}dx
\end{equation}

\vspace{0.5in}

If no infrared cutoff is present ($\lambda=0$), the large $n$
behavior of the $x$ integral is dominated by $x\sim n$ and we have
\begin{equation}
  c_n \sim (\frac{b_0}{2})^{n}n!
\end{equation}
implying a singularity in $\hat{B}(s)$ at $s=\frac{2}{b_0}$. The
{\em absence} of a singularity at the first possible renormalon
location $s=\frac{1}{b_0}$ can be related to the absence of
a local gauge-invariant operator of dimension 2 in QCD (the extra
factor of $l^{2}$ in the numerator of (\ref{eq:IRint}) is a
direct consequence of gauge-invariance). The nonperturbative ambiguity
associated with the singularity at $s=\frac{2}{b_0}$ is of the
form $e^{-\frac{2}{b_0}\alpha(Q^{2})}\sim\frac{1}{Q^{4}}$, exactly
the leading power behavior associated with coefficient functions
of the lowest dimension nontrivial operator $F_{\mu\nu}^{2}$
appearing in the operator product expansion of this current
correlator. 

 With an infrared cutoff present, the renormalon singularity 
disappears, as the large n behavior is dominated by $x\sim\ln{(
\frac{Q^{2}}{\lambda^{2}})}$ and
\begin{equation}\label{eq:IRpresent}
  c_n \sim \ln{(\frac{Q^{2}}{\lambda^{2}})}^{n}
\end{equation}
This power growth means that the contribution from these graphs to
$\hat{B}(s)$ is {\em entire} in $s$. Of course, the Taylor series
generated by the coefficients (\ref{eq:IRpresent}) has a finite
radius of convergence (before the Borel transform). The divergence
occurring when $\lambda$ is decreased to the point where
$\ln{(Q^{2}/\lambda^{2})}\sim\frac{1}{b_0\alpha(Q^{2})}$ is just
the entrance of the Landau singularity into the range of the
momentum integration.

\begin{figure}[htp]
\hbox to \hsize{\hss\psfig{figure=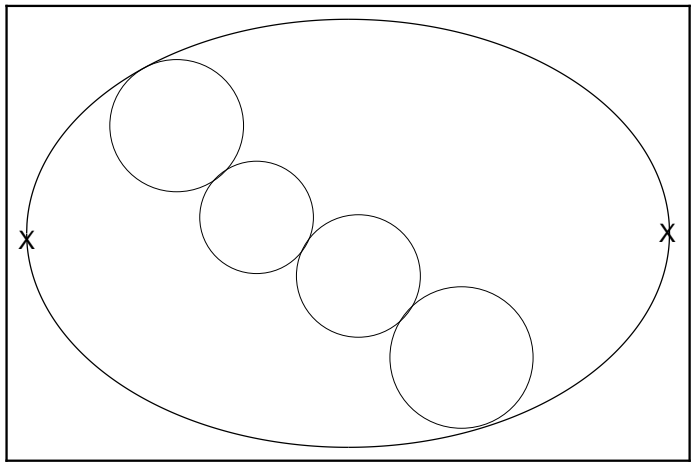,width=0.8\hsize}\hss}
\caption{One of a set of bubble graphs contributing to the leading
perturbative renormalon in the 2D sigma model}
\label{fig:2Dgraph}
\end{figure}

  The preceding discussion applies almost word for word to the
two current correlator (\ref{eq:2curr}) in the 2D sigma model. A
typical set of diagrams giving an infrared renormalon in this 
theory is shown in Fig(\ref{fig:2Dgraph}). Here the current insertions
do not decouple at zero momentum and we have, instead of
(\ref{eq:IRint}),
\begin{equation}
 \int_{\lambda^{2}}dl^{2}\frac{\alpha(Q^{2})}{1+b_0\alpha(Q^{2})
\ln{(\frac{l^{2}}{Q^{2}})}}\sim\sum\alpha^{n+1}(Q^{2})b_0^{n}
\int^{\ln{(\frac{Q^{2}}{\lambda^{2}})}}x^n e^{-x}dx
\end{equation}
If the infrared cutoff is set to zero, the large order behavior
is $c_n\sim b_0^{n}n!$, leading to a singularity in $\hat{B}(s)$
at $s=\frac{1}{b_0}$. This corresponds to a local operator
$|\partial_{\mu}\vec{\phi}|^{2}$ of dimension 2. If
$\lambda \neq 0$, the perturbative renormalon disappears
completely, as the large order growth is power rather than
factorial. 
 
  An important advantage of the 2D sigma model is the availability
of the large $N$ saddle-point technique as a reliable tool for
exhibiting the mechanism of dynamical mass generation in the theory.
It was pointed out previously that dynamical mass generation
occurs even if an infrared cutoff is imposed on the modes of
the elementary $\vec{\phi}$ fields of the theory (recall (\ref{eq:masscut}),
where $m,\Lambda,\lambda$ represent the squared mass, UV and IR
cutoffs resp.). From (\ref{eq:masscut}) it is apparent that even if
$\lambda\neq 0$, the Borel Transform $B(s)$ of the propagator
$\frac{1}{q^{2}+m}$ has simple poles at $s=4\pi n$, just as in
(\ref{eq:Bsprop}). However, we have just seen that if $\lambda\neq 0$,
the usual {\em perturbative} sources of infrared renormalon
singularities in the naive transform $\hat{B}(s)$ disappear!
In fact, the renormalon structure of the full Borel function
$B(s)$ in this case is associated with the appearance of a
zero-momentum condensate for the auxiliary O(N) singlet field
$\rho(x)$ (see (\ref{eq:2DLatt})), which couples to the singlet 
two $\phi$ channel. Even though the individual
$\vec{\phi}$ constituents are restricted to have nonzero 
momentum $k^{2}>\lambda$, {\em pairs} of these quanta can
condense at zero momentum, precisely as in the case of Cooper
pairs in BCS superconductivity. The moral of this example is
clear: the full Borel transform $B(s)$ can have singularities
of purely nonperturbative origin, not associated (at least in
any direct fashion) with the usual IR renormalons of perturbation
theory.

 The preceding discussion applies to the renormalon structure of
the continuum field theory in which the momentum modes are
still continuous, but a sharp infrared cutoff is introduced.
The introduction of a finite lattice imposes an infrared cutoff
uniformly on all fields of the theory, including the condensate
field $\rho$, thereby eliminating the singularities at
$s=4\pi n$ of $B(s)$.

  It was pointed out in section 2 that a singularity in $\tilde{B}(s)$
for the Borel transform of the partition function $Z(f)$ is directly
related to a singularity in the density of configurations in the
defining functional integral for $Z$. Consider the current-current
correlator in quenched QCD (i.e. dropping internal quark loops). 
The UV cutoff will be implemented on a lattice (with a bare coupling
$\alpha\equiv g_0^{2}/4\pi$) and the compact link integrations
written $\int{\cal D}A_{\mu}$ for simplicity. Then (again, representing
the lattice Fourier transform in continuum notation for 
simplicity): 
\begin{eqnarray}\label{eq:PiQCD}
\Pi(Q^{2})&=&N(Q^{2};f)/Z(f) \nonumber \\  
 N(Q^{2};f)&\equiv&\frac{1}{3Q^{2}}\int d^{4}x e^{iq\cdot x}\int{\cal D}A_{\mu}{\rm Tr}
[\gamma^{\mu}\Delta(x;A_{\mu})\gamma_{\mu}\Delta(-x;A_{\mu})]e^{-\frac{1}{\alpha}
S[A]} \nonumber \\
 &=&\int{\cal D}A_{\mu} {\cal P}[A;q]e^{-\frac{1}{\alpha}S[A]}
\end{eqnarray}
where $\Delta(x;A)$ is the quark propagator in the background field $A$.
The Borel discontinuity of $N(Q^{2};f)$ is
\begin{eqnarray}\label{eq:BsPi}
 \tilde{B}_N(s;Q^{2})&=&\int{\cal D}A_{\mu}{\cal P}[A;q]\delta(s-S[A_{\mu}]) \nonumber \\
    &=&\bar{{\cal P}}(s;q)\tilde{B}_Z(s),\;\;\;\;\tilde{B}_Z(s)\equiv\int{\cal D}A_{\mu}\delta(
s-S[A_{\mu}])
\end{eqnarray}
Here $\bar{{\cal P}}(s;q^{2})$ is the average of ${\cal P}[A_{\mu};q]$ over all configurations
with action equal to $s$. $\tilde{B}_Z(s)$ is the fundamental density of configurations in the
theory. As pointed out by t'Hooft in his seminal study of the renormalon structure of QCD
\cite{tHooft}, Borel singularities in the transform of $N(f)$ or $Z(f)$
 will generically transfer, via the 
convolution theorem, to the Borel transform of the connected ratio $N/Z$, barring some miraculous
cancellation. In fact, the location of renormalon singularities (though not their type and
strength) are expected to be substantially independent of the specific Green's function under
consideration: rather, they derive from the basic structure of the functional configuration
space of the theory.

 It has sometimes been asserted that the presence or absence of renormalon singularities 
can be inferred solely from the assumption of nonperturbative validity of the Wilson
operator product expansion. For example, the absence of a perturbative IR renormalon
in $\hat{B}(s)$ at $s=1/b_0$ was related above to the absence of local gauge-invariant
operators of dimension 2. Such a singularity will never be visible in dimensionally
renormalized massless perturbative QCD, where the only mass scale in the theory is
introduced via the regularization procedure in such a way as to result in purely
logarithmic (rather than power) contributions to renormalized quantities. Rather, we
are concerned here with intrinsically nonperturbative contributions to the
coefficient function $C_{1}(x)$ of the identity operator, which is present regardless 
of the subsequent set of nontrivial operators in the theory. Such contributions,
of the form $mx^{2}$ in coordinate space, where $m$ is a dynamically generated
squared mass of the form $\sim e^{-\frac{1}{b_0\alpha}},\;\;\alpha\rightarrow 0$,
inevitably produce a Borel singularity at the {\em first} location $s=\frac{1}{b_0}$.
On the other hand, they 
vanish formally to all orders in a weak coupling expansion, and (in a non-Borel
theory) are {\em not necessarily reconstructible} from conventional sources of factorial
growth in perturbation theory (cf. situation discussed above for 2D sigma model
in presence of an IR cutoff!).

  Finally, let us note that the presumed existence of a mass gap in the theory can be
viewed as prima facie evidence for a universal infrared renormalon singularity at
the first available location. It is very difficult to see how a singularity at
$s=1/b_0$ can be avoided in almost any physical quantity
in  QCD once such a mass gap is assumed, together with
the renormalization group and asymptotic freedom. As the simplest example, consider the 
finite temperature partition function of pure QCD at low temperature $\beta^{-1}$.
Evidently, for large $\beta$, the one glueball sector dominates and (defining
$m\equiv ({\rm glueball-mass})^{2}$)
\begin{eqnarray}\label{eq:Zbeta}
 Z_{\beta}&=&1+V\int\frac{d^{3}p}{(2\pi)^{3}}e^{-\beta\sqrt{(p^{2}+m)}}+..\nonumber \\
 m&=&\Lambda e^{-\frac{1}{b_0\alpha}}
\end{eqnarray}
The higher order terms are suppressed at low temperature, so cannot be expected to 
cancel (except accidentally) any renormalon singularity found in the first nontrivial
term. The Borel transform $B_1(s)$ of the 1-glueball contribution to the free energy
$\ln{Z}/V$ thus becomes
\begin{equation}\label{eq:BZbeta}
  B_1(s)=\int_{0}^{c}dm\;m^{-1-b_0 s}\int\frac{d^{3}p}{(2\pi)^{3}}e^{-\beta\sqrt
{(p^{2}+m)}}
\end{equation}
Integrating by parts to expose the pole at $s=\frac{1}{b_0}$ we find
\begin{equation}
 B_1(s)\sim \frac{1}{1-b_0 s}\int\frac{d^{3}p}{(2\pi)^{3}}\frac{\beta e^{-\beta p}}{2p}
\end{equation}
A similar singularity is expected to arise in all connected Green's functions of the theory
once dynamical mass generation occurs.

   Of course, to settle this contentious issue definitively, it will be necessary to
perform reliable, fully nonperturbative calculations of the generalized Borel function
$B(s)$. The only tool available for doing this in the case of QCD is lattice theory, 
so we must now face directly the issue of extracting renormalon structure of field
theories formulated on a (necessarily) finite space-time lattice.
  
\section{Finding renormalon singularities on a finite lattice} 

  If, as we have argued, weak-coupling perturbation theory is not a reliable
guide to the full renormalon structure in an infrared unstable theory
with dynamical mass generation, it will be necessary to carry out a fully
nonperturbative evaluation of the Borel function $B(s)$ for the relevant
physical quantity. In QCD, this restricts us to a numerical (Monte Carlo)
simulation of the lattice gauge theory. There are two obvious difficulties
which must be overcome. Firstly, the information obtained in such a
calculation is necessarily subject to statistical errors. Consequently,
any technique employed for the extraction of the singularity structure 
of $B(s)$ for Re$(s)>0$ (obtained by analytic continuation from Re$(s)<0$)
must be fairly resistant to the inevitably noisy input information.
Secondly, simulations can only be performed of systems with a finite number
of degrees of freedom, i.e. at finite volume. The Borel Transform $B(s)$ 
(\ref{eq:GBT}) will in this case lack the Type 2 singularities associated
with the condensation of zero momentum modes in the theory. In this section
we shall explain one possible modification of the definition (\ref{eq:GBT})
capable of revealing the precursors of the infinite volume singularity
structure on a finite lattice. 

   As usual, a convenient model for studying the renormalon structure on a 
finite lattice is the O(N) sigma model in 2 dimensions: the large N limit of
the theory is analytically solvable, even on the lattice. We shall use this
model to gain insight in formulating a modified lattice Borel transform.We
begin by pointing out a basic difficulty in the definition (\ref{eq:GBT}) of
the full Borel function $B_L(s)$ when the transform is applied to a lattice system
with a mass gap generated via logarithmic infrared divergencies. To make
the algebra as transparent as possible, consider the toy model of Section 2
where the gap equation is modelled by the discrete sum
\begin{equation}\label{eq:toygap}
\frac{1}{f}=\frac{1}{L}\sum_{n=1}^{L}\frac{1}{m+n/L}\equiv s_L(m)\rightarrow \ln
{(\frac{1+m}{m})},\;\;L\rightarrow\infty
\end{equation}
Here $m$ is the analog of the squared dynamical mass and we take $G(m)=\frac{1}{p^{2}+m}$
as our model of the momentum space propagator, with Borel transform
\begin{eqnarray}\label{eq:toyBT}
B_L(s)&=&\int_{0}^{\infty}e^{s/f}G(m(f))d(\frac{1}{f}),\;\;{\rm Re}(s)<0  \nonumber \\
 &=& -\int_{0}^{\infty}s_L^{\prime}(m)e^{ss_L(m)}G(m)dm
\end{eqnarray}
This integral defines the analytic function $B_L(s)$ for ${\rm Re}(s)<0$. The problem 
is that the $L\rightarrow\infty$ limit does not commute with the desired analytic
continuation to the right-half $s$-plane. This becomes clear if we subject (\ref{eq:toyBT})
to an integration by parts
\begin{equation}\label{eq:BLparts}  
B_L(s)=\frac{1}{s}e^{ss_L(0)}G(0)+\frac{1}{s}\int e^{ss_L(m)}G^{\prime}(m)dm
\end{equation}
For $0<{\rm Re}(s)<1$, the integral in (\ref{eq:BLparts}) is now well-defined
as $L\rightarrow\infty$ (as the integration is convergent for $m\rightarrow 0$),
but $s_L(0)\sim \ln(L)$ causes a power divergence $\propto L^{s}$ if the large volume
limit is taken after the analytic continuation is performed.

  In the case of the sigma model at large $N$, where we have
analytic control of the theory, one may devise the following
cure for this problem, which arises from the deviation at finite $L$
in the integration measure for small $m$ from the infinite volume form,
Namely, one devises a modified Borel transform on the lattice which has a
smooth analytic continuation to ${\rm Re}(s)>0$ at large $L$. In the
$L\rightarrow\infty$ limit, the integration measure goes over to
$m^{-1-s}dm$. Choosing an infrared cutoff $\beta\sim O(1/L)$ (the
precise value is immaterial) we may define an alternative transform of 
the (nonperturbatively determined!) propagator $G(m)$ on the lattice as
\begin{equation}\label{eq:BTmod}
B_{\rm mod}(s)\equiv e^{\eta s}\int_{0}^{\beta}m^{-1-s}G(m-\beta)dm\;\;+B_L(s)
\end{equation}
with $\eta,\beta$ satisfying $\eta=s_L(0)-\ln{(1/\beta)}$, i.e
\begin{equation}\label{eq:constraint}
 e^{s(s_L(0)-\eta)}=\beta^{-s}
\end{equation}
The relation (\ref{eq:constraint}) ensures that the integrand $m^{-s}G(m-\beta)$
in the first term in (\ref{eq:BTmod}) matches (at $m=\beta$) smoothly to the
corresponding lattice determined integrand $e^{ss_L(m)}G(m)$ in $B_L(s)$. The
supplementary contribution generates a term $-\frac{1}{s}e^{\eta s}G(0)\sim L^{s}$
as $L\rightarrow\infty$, exactly cancelling the pathological term in (\ref{eq:BLparts}).
  We have checked that the above procedure leads to a numerically stable detection
procedure for renormalon singularities using the test case of the 2D sigma model. As input to
the Borel transform, the feasibility of the method can be studied by using the 
analytic large N results (instead of actual Monte Carlo data). Statistical noise
can also be introduced by hand to check the robustness of the procedure. 
  The mass gap equation on a finite lattice for the 2D sigma model is
\begin{equation}\label{eq:lattgap}
 \frac{1}{f}=\frac{1}{L^{2}}\sum_{\vec{k}}\frac{1}{d(\vec{k})+m}\equiv \frac{1}{4\pi}s_L(m)
\end{equation}
where $\vec{k}=(\frac{2\pi}{L}k_x,\frac{2\pi}{L}k_y),\;k_x,k_y=1,2,..L$ and
$d(\vec{k})\equiv 4(\sin(\frac{\pi k_x}{L})^{2}+\sin(\frac{\pi k_y}{L})^{2})$. In the
large N limit the momentum space $\vec{\phi}$ propagator is
\begin{equation}\label{eq:Glatt}
 G(f;\vec{q})=\frac{1}{d(\vec{q})+m(f)}
\end{equation}

\begin{figure}[htp]
\hbox to \hsize{\hss\psfig{figure=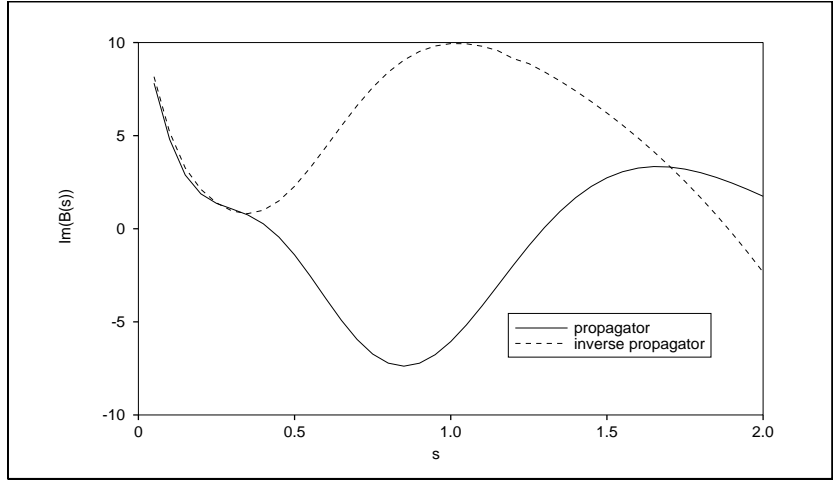,width=0.8\hsize}\hss}
\caption{Borel Transform of propagator, inverse propagator on a
30x30 lattice ($\gamma=0.1,  N_T=10$)}
\label{fig:GinvG}    
\end{figure}

\begin{figure}[htp]
\hbox to \hsize{\hss\psfig{figure=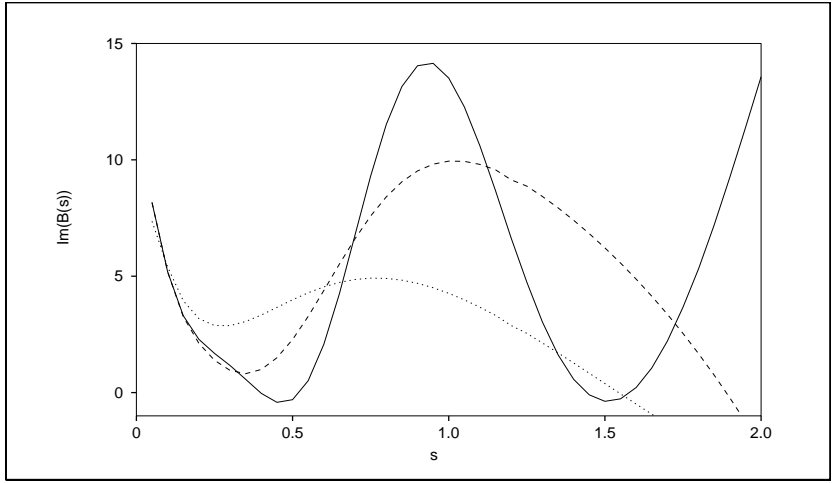,width=0.8\hsize}\hss}
\caption{Borel Transform of  inverse propagator on a
30x30 lattice ($N_T=6,10,14$)}
\label{fig:moreterms}
\end{figure}

In analogy to (\ref{eq:BTmod}), we analytically continue the modified transform (suppressing
for notational simplicity the momentum dependence)
\begin{equation}\label{eq:Blattmod}
 B_{\rm mod}(s)\equiv e^{\eta s}\int_{0}^{\beta}m^{-1-s}G(m(f))dm -\int_{0}^{c}s_L^{\prime}(m)e^{
s\;s_L(m)}G(m)dm
\end{equation}
with $\beta\sim O(1/L^{2})$, $\eta=s_L(0)-\ln(1/\beta)$. The analytic continuation to 
positive real $s$ can be performed in a numerically stable way by writing
\begin{equation}
  s = s_R e^{w}
\end{equation}
and Taylor expanding around $w=i\pi$ (where $B_{\rm mod}(s)$ is analytic). The circle of
convergence has radius $\pi$, as $B_{\rm mod}$ is analytic in the cut plane, and we can
evaluate  ${\rm Im}\; B_{\rm mod}(s_R +i\gamma)$ by summing the series to some finite order
$N_T$ at $w \sim i\gamma$. The result for $N_T=10,\gamma=0.1$ 
on a 30x30 lattice is shown in Fig(\ref{fig:GinvG})
for the Borel transform of both $G$ (with $d(\vec{q})=1$) and the inverse propagator $G^{-1}$.
The effect of keeping more terms in the Taylor expansion is shown in Fig(\ref{fig:moreterms})
for the Borel Transform of $G^{-1}$. The peak at the first renormalon becomes sharper as $N_T$
increases. However, the results become (for fixed $\gamma$) less accurate as $s_R$ increases,
which requires $w$ to approach the edge of the circle of convergence. 

\begin{figure}[htp]
\hbox to \hsize{\hss\psfig{figure=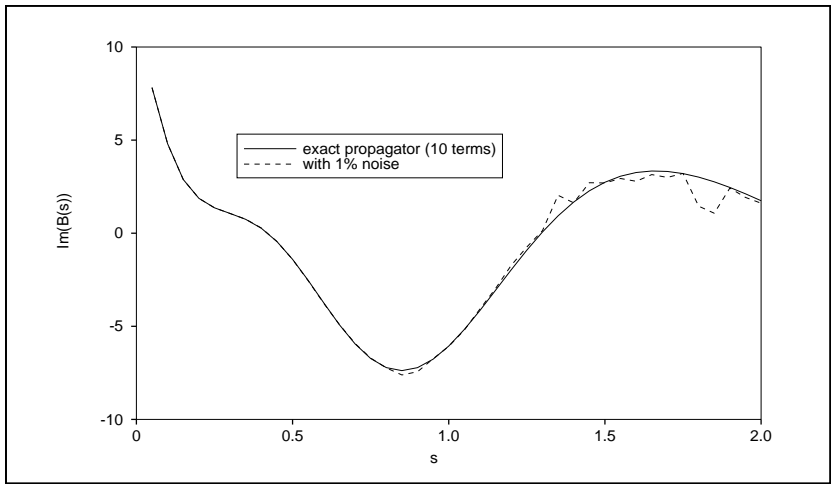,width=0.8\hsize}\hss}
\caption{Borel Transform of noisy propagator on a
30x30 lattice ($N_T=10$)}
\label{fig:noisy}    
\end{figure}

Finally, the sensitivity of the continuation procedure
to a random superimposed 1\% statistical noise on $G(m)$ (well within the reach of Monte Carlo
simulations, for example) is shown in Fig(\ref{fig:noisy}). The first renormalon is still
clearly visible at $s\sim 1$, with the effects of the noise becoming serious only for
$s>1.5$. In all of these cases, we should emphasize that the first term in (\ref{eq:Blattmod})
is absolutely essential to obtain stable results from the analytic continuation.

\section{Summary and Discussion}

  In this paper we have used the nonlinear sigma model as a convenient analytical
platform for examining the interrelationship of various versions of the Borel
transform. Infrared unstable theories (like QCD or the sigma model) with 
dynamical mass generation are inevitably non-Borel summable, and a precise Borel
reconstruction of such theories is only possible starting from the generalized
Borel transform $B(s)$, defined as the Laplace transform with respect to the inverse
bare coupling of the relevant amplitude.(The complex nonperturbative
behavior of such theories is entirely
connected with the infrared structure, so it suffices to work throughout with
an ultraviolet cutoff: in the case of QCD, this implies a lattice formulation.)

  Unfortunately, the conventional definition of the Borel transform in terms of
the formal perturbative series (for which we have used the notation $\hat{B}(s)$)
does not lead to a precise reconstruction theorem in the nonBorel case. This
function extends (in the sense of having identical asymptotic expansion for
small $s$) to a function $\tilde{B}(s)$, the discontinuity of the generalized
transform $B(s)$ introduced above, which does however yield such a reconstruction.
The essential difference between the two functions can be seen explicitly in the
large $N$ limit of the nonlinear sigma model, where an infrared cutoff in the
elementary modes of the theory  can be seen to eliminate the conventional 
infrared renormalons of perturbation theory, while leaving the singularities in
$B(s)$ intact. 

  The necessity for a fully nonperturbative approach to the generalized Borel
function $B(s)$ leads us to the consideration of simulations on a finite lattice,
which is often the only tool available for reliable nonperturbative 
calculation. The Borel singularities on the positive $s$ axis have been 
seen to fall into two types. Type 1 singularities are present (as actual
singularities) already in the finite volume theory: however they are numerically
camouflaged by perturbative prefactors in disconnected Green's functions.
Finding such singularities on a finite lattice will therefore require 
removing disconnected vacuum contributions (cf. Section 4). Type 2 singularities
(the usual IR renormalons fall into this category) are strictly speaking
only present in the infinite volume limit. Moreover, the interchange of
analytic continuation to Re($s)>0$ and the infinite volume limit is 
not uniform, requiring a modification (discussed in Section 6) of the
definition of the Borel Transform on a finite lattice in order to allow
detection of the renormalon precursors. Such a modification indeed allows
us to extract the location of renormalon singularities for the analytically
tractable test case of the sigma model at large $N$. The application to 
situations (say, the sigma model with $N=3$, or quenched QCD)
 where Monte Carlo methods must be employed to
compute the nonperturbative structure of the theory is presently under study.

\newpage

\section{Acknowledgements}

 A.D. gratefully acknowledges many useful and enlightening conversations
on the role of renormalons with A.H. Mueller.

\end{document}